\documentclass{article}
\usepackage{spconf,graphicx}
\usepackage[caption=false]{subfig}
\usepackage{amsmath, mathtools, stmaryrd, amssymb}
\usepackage[nolist]{acronym}
\graphicspath{{../figs/}}
\usepackage[noadjust]{cite}	
\usepackage{url}

\def\y{{\mathbf y}}

\def\w{{\mathbf w}}
\def\z{{\mathbf z}}

\newcommand\ub[1]{\underbar{$#1$}}

\title{Distributed speech separation in spatially unconstrained microphone arrays}
\twoauthors
{Nicolas Furnon\thanks{This work was made with the support of the French National Research Agency, in the framework of the  project DiSCogs “Distant speech communication with heterogeneous unconstrained microphone arrays” (ANR-17-CE23-0026-01). Experiments presented in this paper were partially carried out using the Grid5000 testbed, supported by a scientific interest group hosted by Inria and including CNRS, RENATER and several Universities as well as other organizations (see https://www.grid5000).},
	Romain Serizel, Irina Illina}
{Universit{\'{e}} de Lorraine, CNRS, Inria, Loria \\ F-54000 Nancy, France \\ \{firstname.lastname\}@loria.fr}
{Slim Essid}
{LTCI, T\'el\'ecom Paris, \\ Institut Polytechnique de Paris, \\Palaiseau, France \\
	slim.essid@telecom-paristech.fr}
\begin{document}
\ninept
\maketitle
\begin{acronym}
\acro{ds}[DSB]{delay-and-sum beamformer}
\acro{mpdr}[MPDR]{minimum power distortionless response beamformer}
\acro{mvdr}[MVDR]{minimum variance distortionless response beamformer}
\acro{lcmp}[LCMP]{linearly constrained minimum power beamformer}
\acro{lcmv}[LCMV]{linearly constrained minimum variance beamformer}
\acro{mwf}[MWF]{multichannel Wiener filter}
\acro{sdw}[SDW-MWF]{speech distortion weighted multichannel Wiener filter}
\acro{mvdr}[MVDR]{minimum variance distortionless response}
\acro{gevd}[GEVD]{generalized eigenvalue decomposition}
\acro{nmf}[NMF-MWF]{non-negative matrix factorization}
\acro{stft}[STFT]{short-time Fourier transform}
\acroplural{stft}[STFTs]{short-time Fourier transforms}
\acro{tf}[TF]{time-frequency}
\acro{vad}[VAD]{voice activity detector}
\acroplural{vad}[VADs]{voice activity detectors}
\acro{danse}[DANSE]{distributed adaptive node-specific signal estimation}
\acro{mse}[MSE]{mean squared error}
\acro{wasn}[WASN]{wireless acoustic sensor network}
\acroplural{wasn}[WASNs]{wireless acoustic sensor networks}
\acro{doa}[DOA]{direction of arrival}
\acroplural{doa}[DOAs]{directions of arrival}
\acro{irm}[IRM]{ideal ratio mask}
\acroplural{irm}[IRMs]{ideal ratio masks}
\acro{ibm}[IBM]{ideal binary mask}
\acro{dnn}[DNN]{deep neural network}
\acroplural{dnn}[DNNs]{deep neural networks}
\acro{nn}[NN]{neural network}
\acroplural{nn}[NNs]{neural networks}
\acro{lstm}[LSTM]{long short-term memory}
\acro{cnn}[CDNN]{convolutional neural network}
\acroplural{cnn}[CNNs]{convolutional neural networks}
\acro{gru}[GRU]{gated recurrent unit}
\acro{crnn}[CRNN]{convolutional recurrent neural network}
\acro{rnn}[RNN]{recurrent neural network}
\acroplural{rnn}[RNNs]{recurrent neural networks}
\acro{rir}[RIR]{room impulse response}
\acroplural{rir}[RIRs]{room impulse responses}
\acro{ssn}[SSN]{speech shaped noise}
\acro{snr}[SNR]{signal to noise ratio}
\acroplural{snr}[SNRs]{signal to noise ratios}
\acro{sar}[SAR]{source to artifacts ratio}
\acro{sir}[SIR]{source to interferences ratio}
\acroplural{sir}[SIRs]{source to interferences ratios}
\acro{sdr}[SDR]{source to distortion ratio}
\acro{sisdr}[SI-SDR]{scale-invariant signal to distortion ratio}
\acro{stoi}[STOI]{short-time objective intelligibility}

\end{acronym}

\begin{abstract}
Speech separation with several speakers is a challenging task because of the non-stationarity of the speech and the strong signal similarity between interferent sources. Current state-of-the-art solutions can separate well the different sources using sophisticated deep neural networks which are very tedious to train. When several microphones are available, spatial information can be exploited to design much simpler algorithms to discriminate speakers. We propose a distributed algorithm that can process spatial information in a spatially unconstrained microphone array. The algorithm relies on a convolutional recurrent neural network that can exploit the signal diversity from the distributed nodes. In a typical case of a meeting room, this algorithm can capture an estimate of each source in a first step and propagate it over the microphone array in order to increase the separation performance in a second step. We show that this approach performs even better when the number of sources and nodes increases. We also study the influence of a mismatch in the number of sources between the training and testing conditions.

\end{abstract}
\begin{keywords}
Speech separation, microphone arrays, distributed processing.
\end{keywords}
\section{Introduction}\label{sec:intro}
Speech separation aims at extracting the speech signals of each speaker in a noisy mixture. It has many applications, for example in automatic speech recognition \cite{Barker2018}, hearing aids \cite{Kokkinakis2008} or music processing \cite{Demir2012}. In recent years, \ac{dnn}-based solutions have replaced model-based approaches because of the great progress they enabled \cite{Erdogan2015, Hershey2016, Luo2019, Zhang2020, Zeghidour2020}. However, most of these \ac{dnn}-based solutions are developed in ``clean'' contexts, where the speech signals are not corrupted by noise or by reverberation, which makes them quite unrealistic for real-life applications. Besides, the recent trend shows that state-of-the-art results are mostly achieved with complex \acp{dnn} \cite{Luo2019, Zhang2020, Zeghidour2020, Chen2020}, with very high model sizes \cite{Luo2020b}. They also operate on single-channel data, and hence neglect the spatial information which could be accessible in everyday life scenarios, where the majority of recording devices are embedded with multiple microphones. Some solutions have been designed to address the case of multichannel scenarios \cite{Gu2019, Wang2020} or reverberant conditions \cite{Delfarah2019}, but they result in even more complex \acp{dnn} than those used in the single-channel context. In everyday life scenarios, where several people speak in presence of their personal devices, these solutions are not (yet) applicable because the computational requirements are too demanding for the memory, computing and energy capacities of the devices.

One way to reduce the computational cost of the \ac{dnn}-based methodologies while exploiting spatial information is to use ad-hoc microphone arrays and to distribute the processing over all the devices of the array. In a previous article, we introduced a solution that proved to efficiently process multichannel data in a distributed microphone array in the context of speech enhancement \cite{Furnon2020a}. This approach was based on a two-step version of the \ac{danse} algorithm by Bertrand and Moonen, where so-called compressed signals are sent among the devices \cite{Bertrand2010a}. It achieved, with a simple \ac{crnn}, competitive results compared to a filter computed from oracle \ac{tf} masks. In an extended study, we have shown that exchanging the estimates of both the target and the noise across the devices could potentially increase the final performance \cite{Furnon2020b}. 

In this paper, we show that this solution can be adapted to speech separation in a typical use case of a meeting, where all the sources must be estimated in a reverberant but noise-free environment. We redesign the solution such that each device sends an estimate of a different source in a first step. This way, in a second step, each device has access to proper insight into the whole acoustic scene and can separate the signals more accurately. We analyse the performance over the number of sources present and devices available.

This paper is organised as follows. We formulate the problem in Section \ref{sec:pb_formulation} and detail our solution in Section \ref{sec:contribution}. The experimental setup on which we test the solution is described in Section \ref{sec:setup}. The results are given in Section \ref{sec:results}. We conclude the paper in Section \ref{sec:conclusion}.

\section{Problem formulation}
\label{sec:pb_formulation}
We consider a scenario where $N$ speakers $\{s_n\}_{n=1..N}$ are recorded by $K$ devices, thereafter called nodes. Each node $k$ contains $M_k$ microphones. We assume that no noise is present, so the $m$-th microphone of the $k$-th node records the following signal at a time \mbox{stamp $t$}:
\begin{align*}
y_{m_k}(t) &= \sum_{n=1}^N c_{n,m_k}(t) * s_n(t) \\
~ &=  \sum_{n=1}^N \hat{s}_{n,m_k}(t)
\end{align*}
where $*$ stands for the convolution operator and $c_{n,m_k}$ is the \ac{rir} from the source $n$ to the microphone $m_k$ such that $\hat{s}_{n,m_k} = c_{n,m_k} * s_n$ is the reverberated image of the source $n$ captured by the \mbox{$m_k$-th} microphone. Speech separation aims at recovering all the speech signals $\{\hat{s}_{n,\mu^{(n)}}\}_{n=1..N}$ where $\mu^{(n)}$ is the reference microphone for source $s_{n}$.

In the \ac{stft} domain, under the narrowband approximation, we can write:
\begin{equation*}
	\ub{y}_{m_k}(f, \tau) = \sum_{n=1}^N \hat{\ub{s}}_{n,m_k}(f, \tau)
\end{equation*}
where $f$ is the frequency index and $\tau$ the frame index, and where $\ub{y}$ and $\hat{\ub{s}}$ are the \acp{stft} of $y$ and $\hat{s}$ respectively. In the sequel, for the sake of conciseness, we will omit the time and frequency indexes. The under bar $\ub{\cdot}$ indicates a signal in the \ac{stft} domain. 

In the context of ad-hoc microphone arrays, we can gather all the microphones of one node into the column vector:
\[
\ub{\y}_k = \left[ \ub{y}_{1_k}, \dots, \ub{y}_{{M_k}_k} \right]^T\,,
\]
and the vectors of the signals of all nodes into another vector:
\[
\ub{\y} = \left[ \ub{\y}_{1}^T, \dots, \ub{\y}_{K}^T \right]^T\,.
\]

\section{Distributed speech  separation algorithm}\label{sec:contribution}
\begin{figure}
	\centering
	\includegraphics[width=\linewidth]{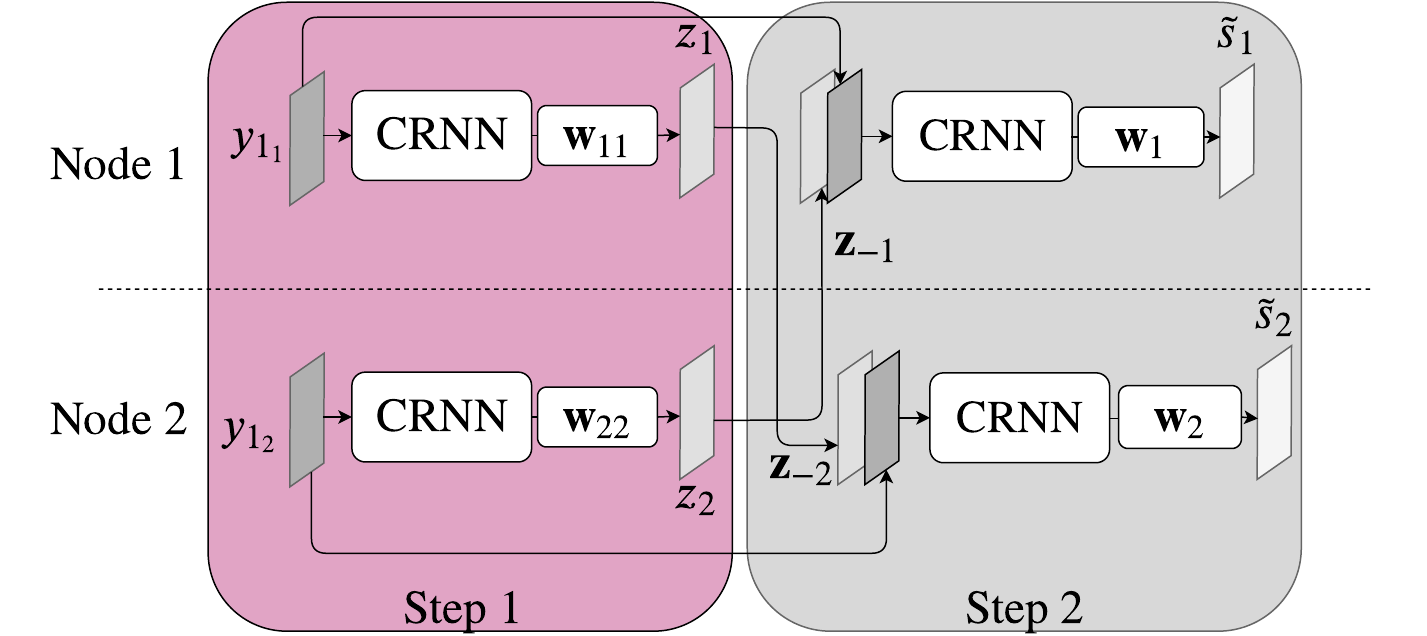}
	\caption{Illustration of our solution in a two-node context. "CRNN" refers to convolutional recurrent neural network.}
	\label{fig:tango}
\end{figure}

To recover the separated signals, we apply a two-step algorithm derived from the \ac{danse} algorithm introduced by Bertrand and Moonen \cite{Bertrand2010a}, and illustrated in Figure \ref{fig:tango} for two nodes. In the first step, at node $k$, the local signals $\ub{\y}_k$ are pre-filtered by a local \ac{mwf} $\w_{kk}$ minimising the \ac{mse} between a single desired speech source $\hat{s}_{k,\mu^{(k)}}$ and the filtered signal:
\begin{equation}\label{eq:cost_wkk}
	\w_{kk} = \mathrm{arg}\min_{\w} \mathbb{E}\{|\hat{s}_{k,\mu^{(k)}} - \w^H\ub{\y}_k|^2\}\,,
\end{equation}
where $\cdot^H$ is the Hermitian transpose operator.
The solution to Equation~\eqref{eq:cost_wkk} yields a so-called compressed signal \mbox{$\ub{z}_k = \w_{kk}^H\ub{\y}_k$} which is a first estimate of the source and sent to all the other nodes. Similarly, the node $k$ receives $K-1$ compressed signals $\ub{\z}_{-k}$:
\[
\ub{\z}_{-k} = \left[ \ub{z}_{1}, \dots, \ub{z}_{k-1}, \ub{z}_{k+1}, \dots, \ub{z}_K \right]^T\,,
\]
 which are stacked with its local signals into:
\[
\ub{\tilde{\y}}_k = \left[ \ub{\y}_{k}^T,~\ub{\z}_{-k}^T \right]^T\,.
\]
In the second step, a second \ac{mwf} $\w_k$ is applied on $\ub{\tilde{\y}}_k$ to finally estimate the source $\hat{s}_{k,\mu^{(k)}}$:
\begin{equation}\label{eq:out_wk}
	\ub{\tilde{s}}_{\mu_k} = \w_k^H\ub{\tilde{\y}}_k\,.
\end{equation}

The advantages of this algorithm are twofold. First, it removes the dependency on a central node that would gather all the signals of all nodes. Second, each node has access to the spatial information of the whole microphone array, but in a condensed and pre-filtered form as each node sends only the compressed signal. 

If the number of nodes is at least equal to the number of speakers,  at the scale of one node $k$, the speech separation problem can be viewed as a speech enhancement problem in the presence of interfering speakers, where the target source is $\hat{s}_{k,\mu^{(k)}}$.
Under the assumption that the sources are uncorrelated, the solution to Equation~\eqref{eq:cost_wkk} is given by:
\begin{equation}\label{eq:mwf_wkk}
	\w_{kk} = \mathbf{R}_{y_k}^{-1}\mathbf{R}_{{s}_{k}}\mathbf{e}_1\,.
\end{equation}
$\mathbf{R}_{{y}_k}$ is the spatial covariance matrix of the mixture $\ub{\y}_k$; $\mathbf{R}_{s_{k}}$ is the spatial covariance matrix of the target signal and we have $\mathbf{e}_1~=~[1, 0, \dots, 0]$. The filter in Equation~\eqref{eq:out_wk} can be obtained in a similar way by replacing the covariance matrix of $\ub{\y}_k$ by the covariance matrix of $\ub{\tilde{\y}}_k$ and by computing the target covariance matrix out of the target components of $\ub{\tilde{\y}}_k$.

These different covariance matrices are computed from signals estimated with a \ac{tf} mask applied on the mixture. We use one common mask for all the signals of one node, as we noticed that taking a specific mask for the compressed signals does not influence the final performance \cite{Furnon2020b}. In the first step, the mask is estimated by a single-node \ac{crnn}, which predicts the mask from a single (local) mixture signal of the node it operates on. In the second step, the \ac{tf} mask is estimated by a multi-node \ac{crnn}, which predicts the masks from the local signal together with the compressed signals sent by all the other nodes \cite{Furnon2020b}.

\begin{figure}
	\centering
	\includegraphics[width=\linewidth]{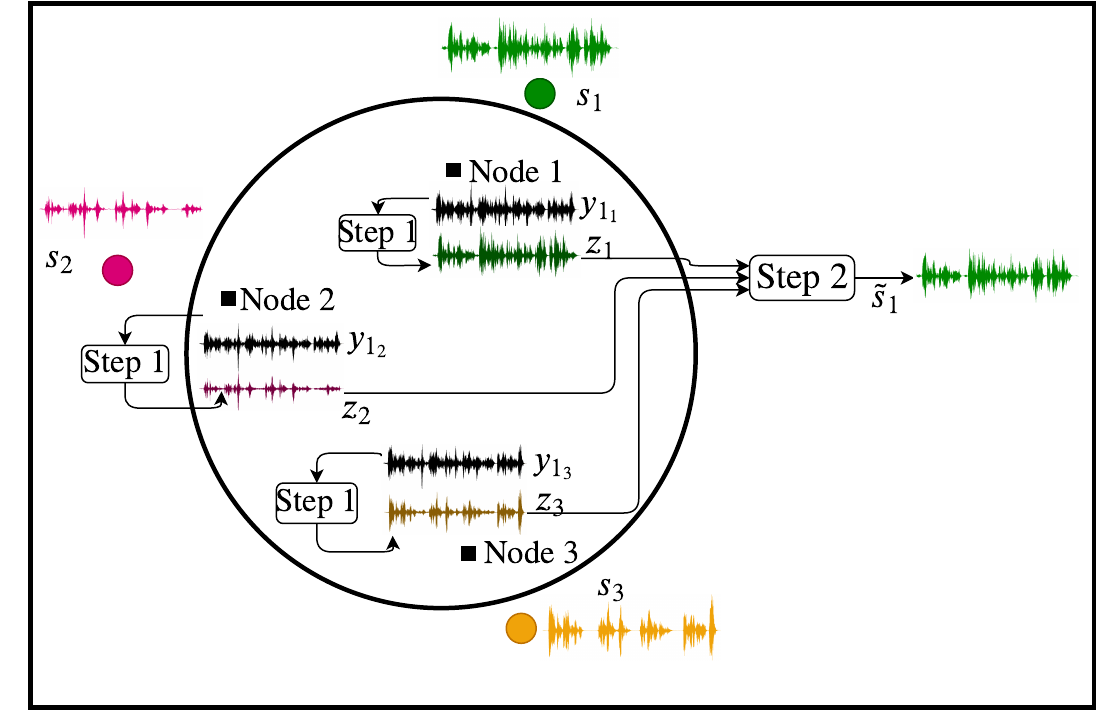}
	\caption{Illustration of our solution in a three-node context, focusing on the first node for the second step.}
	\label{fig:meetit}
\end{figure}

As a consequence of this methodology, in a setup where \mbox{$K \ge N$}, i.e. where there are at least as many nodes as sources, each node can automatically estimate a different source and send a different compressed signal to the other nodes (see Figure~\ref{fig:meetit}). That way, at the second step, each node has an estimate of all the sources, which helps the multi-node neural network to better predict the \ac{tf} mask.

\section{Setup}\label{sec:setup}
\subsection{Dataset}\label{subsec:dataset}
\begin{figure}
	\centering
	\includegraphics[width=.8\linewidth]{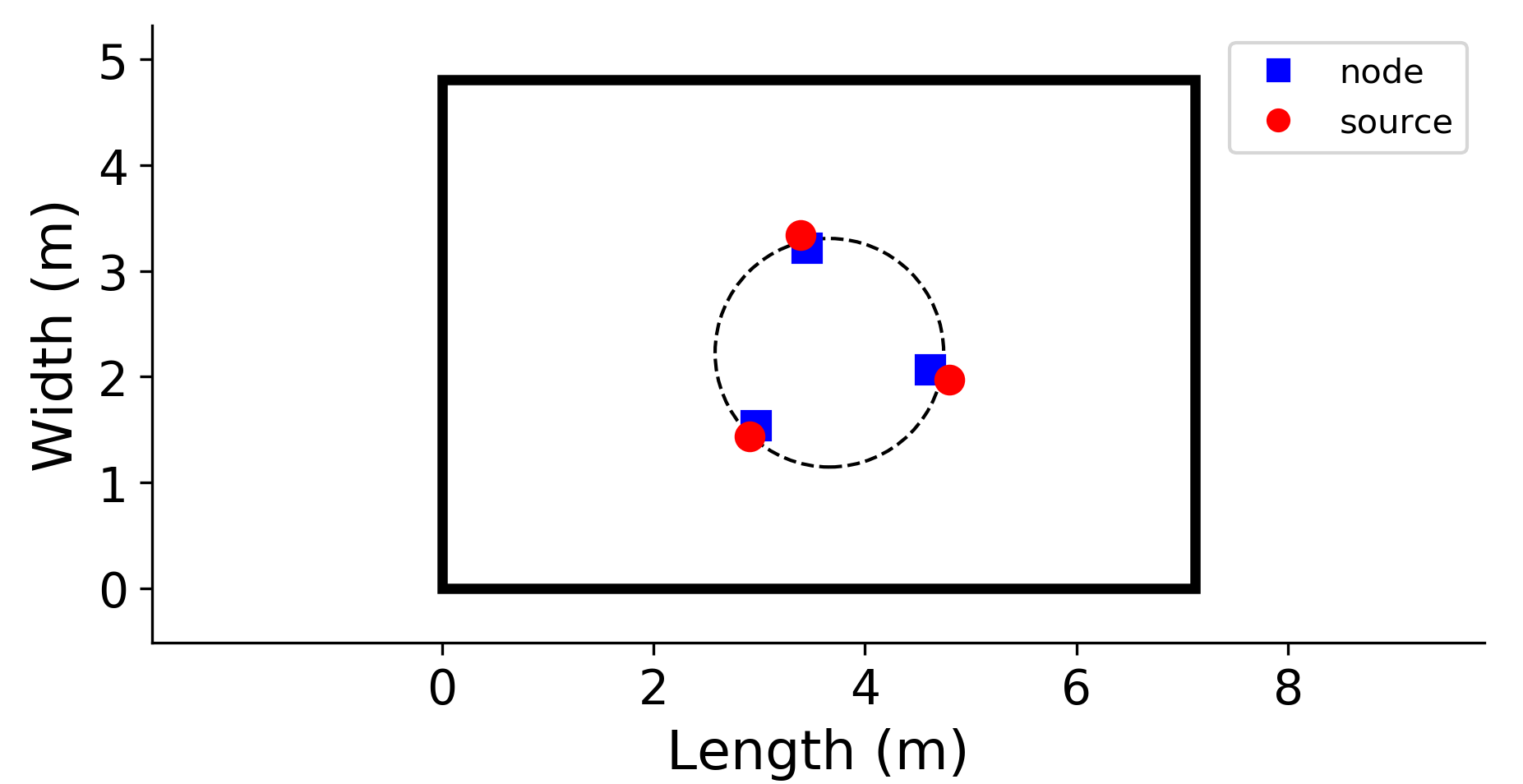}
	\caption{Illustration of one configuration with three sources (hence three nodes).}
	\label{fig:ex_dataset}
\end{figure}

We simulate a typical situation of a meeting, where $N$ persons are talking around a round table, each of the speakers having laid their smartphone, tablet or computer in front of them on the table. We assume that all the nodes have $M_k~=~4$ microphones. The shoebox-like room has a random length, width and height uniformly drawn in [3,~9]~meters, [3,~7]~meters and [2.5,~3]~meters, respectively. The table has a radius randomly drawn between 0.3~m and 2.5~m. The height of the table is randomly selected between 0.8~m and 0.9~m. The sources are evenly placed around the table, so that the angle between two sources is equal for all the pairs of sources. Their distance to the table edge is randomly selected between 0~cm and 50~cm, and their height between 1.15~m and 1.80~m, as if people were sitting or standing close to the table. The reverberation time is randomly selected between 300~ms and 600~ms. The level of all the sources is set to the same power. The \acp{rir} of the room with $N$ equal to 2, 3 and 4 are computed with Pyroomacoustics \cite{Scheibler2018}. The acoustical effect of the table is not simulated. An example with $N=3$ is shown in Figure~\ref{fig:ex_dataset}.

All the speech files are from LibriSpeech \cite{Panayotov2015}. The repartition of the \texttt{train-clean-360}, \texttt{dev-clean} and \texttt{test-clean} subfolders is kept for our split between training, validation and test datasets. Within a mixture, the different speech signals fully overlap in time. We created around 30~hours of training data, 3~hours of validation data and 3~hours of test data.\footnote{A Python implementation to generate the dataset is available at \url{https://github.com/nfurnon/disco/tree/master/dataset_generation/gen_meetit}}

\subsection{Experimental settings}\label{subsec:settings}
All the signals are sampled at a frequency of 16~kHz. The \ac{stft} is computed with a Hanning window of 32~ms with an overlap of 16~ms. The same \ac{crnn} model as the one used in our previous experiments is used \cite{Furnon2020b}. It is made of three convolutional layers, followed by a recurrent layer and a fully-connected layer. The convolutional layers have 32, 64 and 64 filters respectively, with kernel size $3 \times 3$ and stride $1 \times 1$. The recurrent layer is a 256-unit gated recurrent unit, and the activation function of the fully-connected layer is a sigmoid. The network was trained with the RMSprop optimizer \cite{rmsprop}. The input of the model are \ac{stft} windows of 21 frames and the ground truth targetted are the corresponding frames of the \ac{irm}.

\section{Results}\label{sec:results}
We compare four methods in terms of \ac{sisdr} \cite{LeRoux2019}. The first method uses \acp{irm} to compute the signal statistics. In the scenarios we designed, unless a source is removed, node $k$ is always in front of the \mbox{$k$-th} source. At the scale of this node, the speech separation problem is a speech enhancement problem where all the sources \mbox{$j\ne k$} sum to the noise component. Hence we can compute the \ac{irm} at node $k$ as:
\begin{equation}\label{eq:irm}
\text{IRM}_k = \frac{|\ub{\hat{s}}_{k, \mu^{(k)}}|}{|\ub{\hat{s}}_{k, \mu^{(k)}}| + |\ub{n}_{k, \mu^{(k)}}|}
\end{equation}
where $\mu^{(k)}$ is the reference microphone for source $s_{k}$ and \[\ub{n}_{k, \mu^{(k)}} = \sum_{j\ne k}\ub{\hat{s}}_{j, \mu^{(k)}}\,.\]
This oracle method is denoted ``IRM" in the legends of the following figures. The second method, denoted ``MN", is our multi-node solution where the \ac{crnn} sees at the second step the compressed signals to predict the masks. The third method, denoted ``SN", is the single-node solution where the same \ac{crnn} sees only the local signal to predict the masks at both filtering steps. The fourth method, denoted ``MWF", is a \ac{mwf} applied on each node without exchanging the signals. We analyse the behaviour of these methods when the number of sources and nodes varies.

\subsection{Performance with an equal number of sources}
\begin{figure}
	\centering
	\includegraphics[width=.6\linewidth]{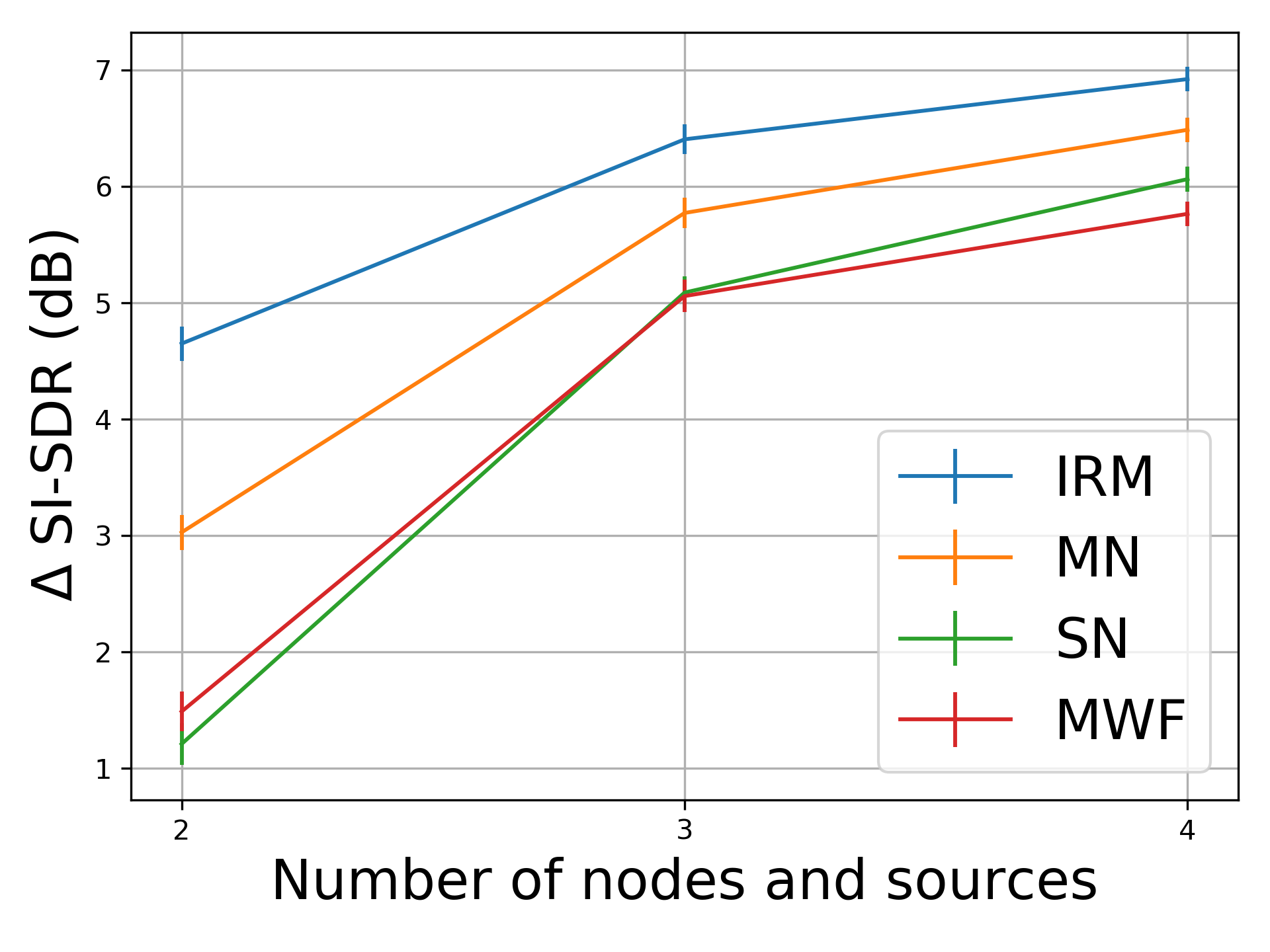}
	\caption{Speech separation performance of the different methods when the number of sources and nodes is equal. The bars correspond to the 95\% confidence interval.}
	\label{fig:equal}
\end{figure}

In this section, the four methods are compared in the scenarios where the number of sources and nodes is equal. The performance in terms of \ac{sisdr} are reported in Figure~\ref{fig:equal}. First, although the single-node solution and the \ac{mwf} do not differ much, the single-node distibuted processing shows a significant improvement over the \ac{mwf} when the number of nodes increases. This shows that exploiting the spatial information conveyed by the whole microphone array helps improving the separation performance. Besides, the multi-node solution significantly outperforms both the single-node solution and the \ac{mwf}. This shows that the compressed signals are useful not only for the beamforming but also for the mask prediction. Finally, the \mbox{$\Delta$SI-SDR} increases when the number of nodes and sources increases, even if the task gets more challenging, and achieves less than 0.5~dB worse than the oracle performance. This is because the output performance remains constant while the input \ac{sisdr} decreases. This shows the robustness of our solution to spatial diversity.
\subsection{Performance in over-determined cases}
\begin{figure}[htb]
	\begin{minipage}[b]{.48\linewidth}
		\centering
		\centerline{\includegraphics[width=5cm]{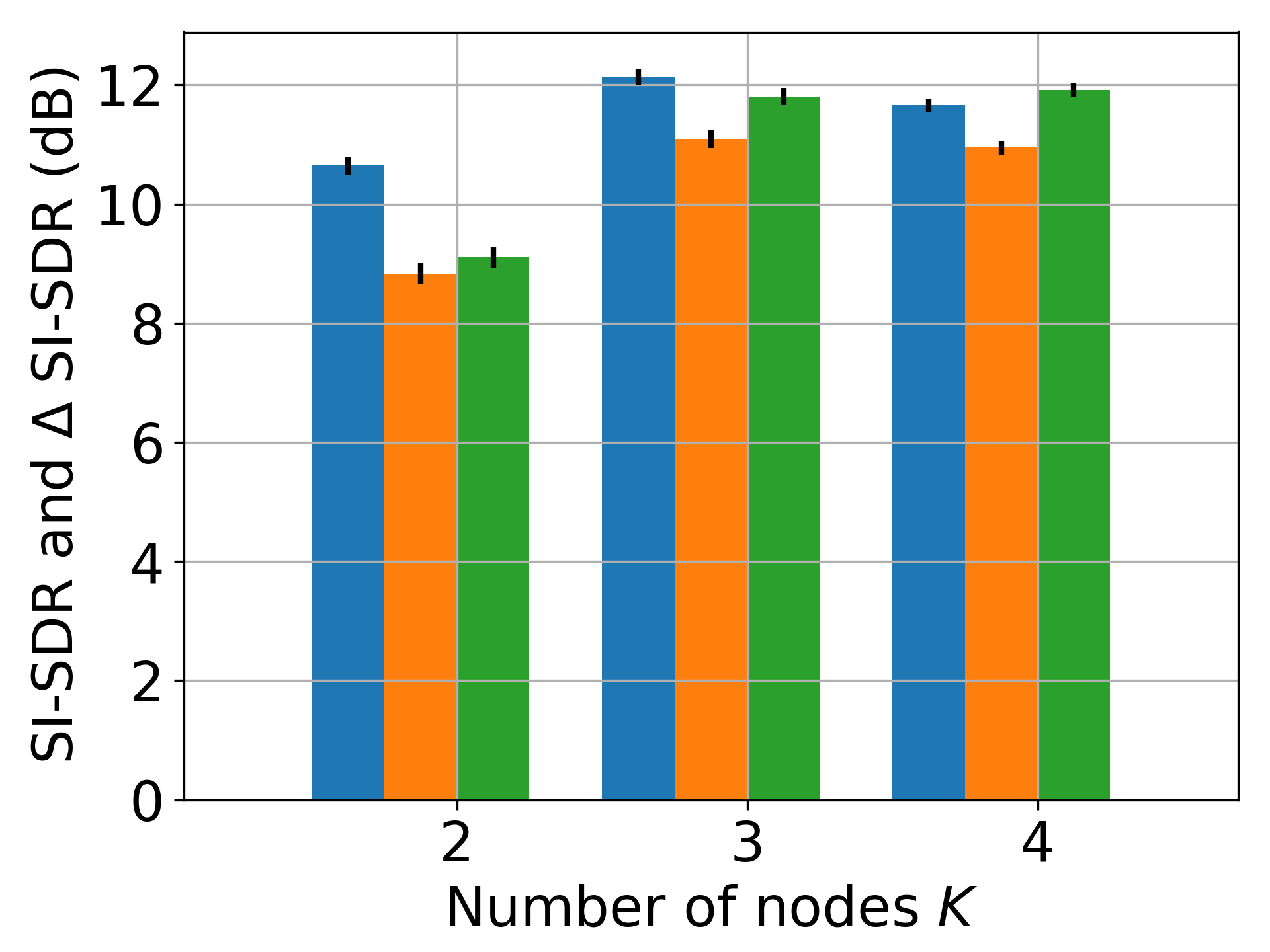}}
		\centerline{(a)}\medskip
	\end{minipage}
	\hfill
	\begin{minipage}[b]{.48\linewidth}
		\centering
		\centerline{\includegraphics[width=5cm]{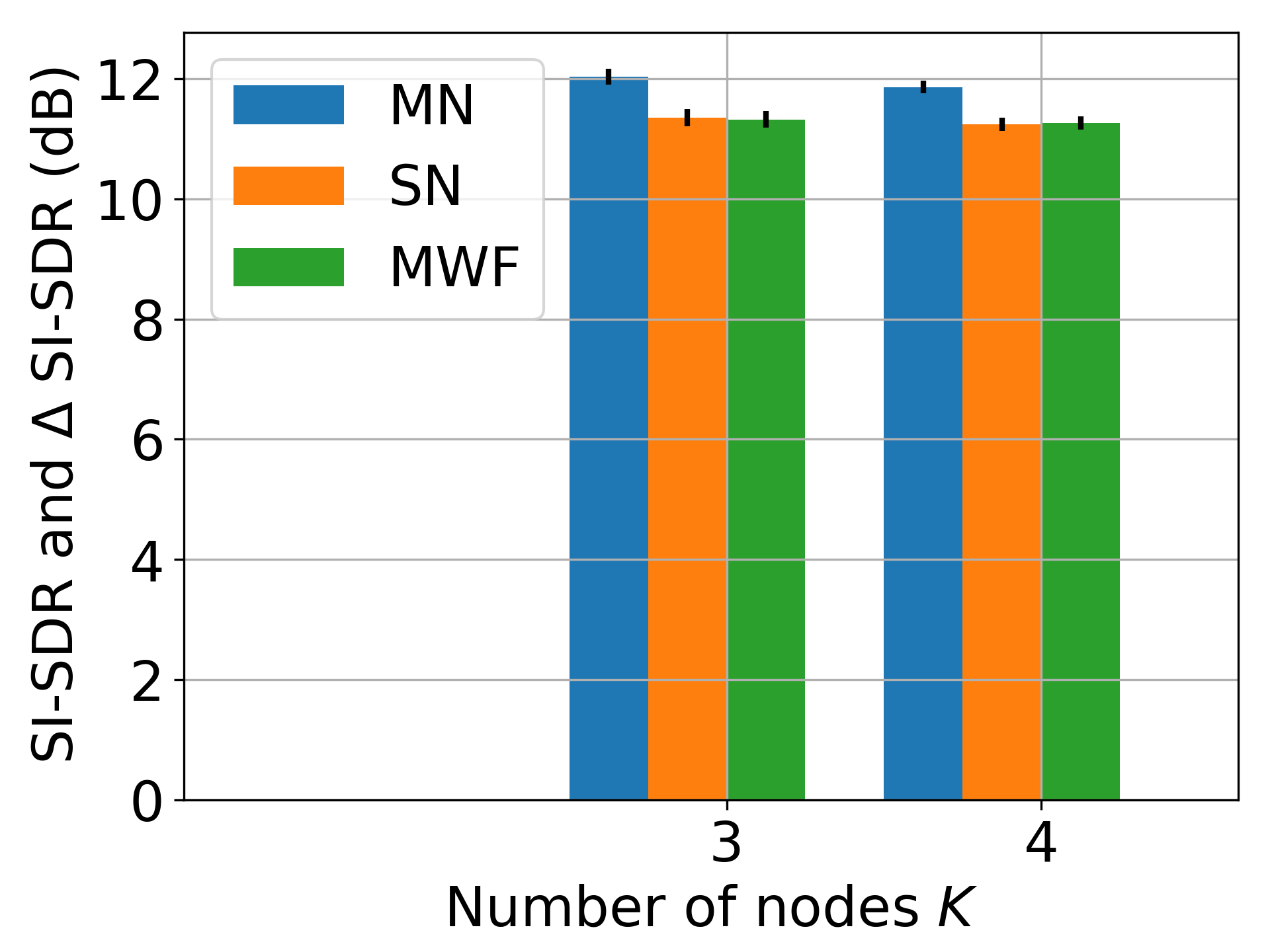}}
		\centerline{(b)}\medskip
	\end{minipage}
	\caption{Speech separation performance in over-determined cases with (a) $N=2$ sources and (b) $N=3$ sources.}
	\label{fig:over}
\end{figure}

In this section, we analyse the performance of the previous methods in the case where the number of nodes is greater than the number of sources. This could happen in a real situation, for example if a person gets out of the room while leaving their phone on the table. In such a scenario, each node estimates the source in front of which it is placed, which means that the node left without speaker has no target speaker. At this node, the target and noise components of Equation~\eqref{eq:irm} are not defined, and it is not obvious to determine the compressed signal to send with the \ac{irm} method. Because of this, we omit the results obtained with the \ac{irm} in this section. The results of the three other methods are reported in Figure~\ref{fig:over} where we recall the performance of the equally-determined case ($N=K=2$ and $N=K=3$) for an easier comparison.

In over-determined cases as well, the multi-node solution outperforms almost always the other two methods. An interesting exception can be noticed when there are two more nodes than sources (\mbox{$N=2$}, \mbox{$K=4$}). We observed that the nodes placed in front of no source estimate completely silent compressed signals, because the masks predicted by the single-node \ac{crnn} are close to 0 in almost all \ac{tf} bins. This means that the filter at the second step is applied on silent signals, which degrades the final performance. Since the multi-node solution still outperforms the single-node solution, the silent compressed signals do not degrade the mask prediction. However, they degrade the distributed beamforming output. This effect is not dominant when there is only one more node than the number of sources, which means that our solution is robust to a source mismatch between training and testing, but only to a limited extent. One solution to cope with this performance drop, could be to automatically ignore the recording devices which start to send silent signals.

Finally, the performance consistently increases for all methods when the number of nodes increases from \mbox{$K=2$} to \mbox{$K=3$} with two sources. This shows that the neural networks trained with more sources (so on harder conditions) perform better than those trained with a lesser number of sources.
\subsection{Performance in under-determined cases}
\begin{figure}[h!]
	\centering
	\includegraphics[width=5cm]{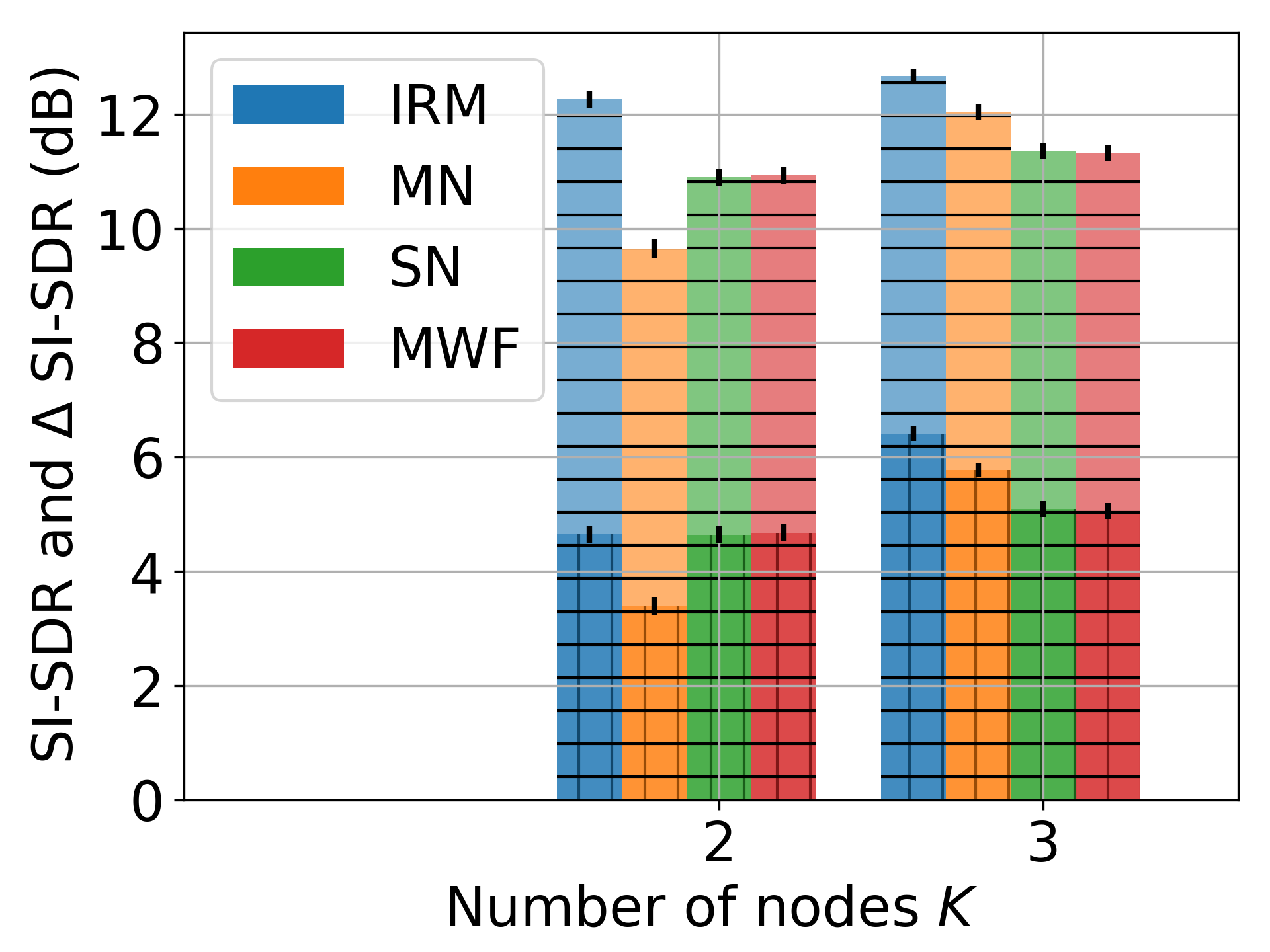}
	\caption{Speech separation performance in an under-determined case with \mbox{$N=3$} sources. The lighter bars with horizontal hatches correspond to the SI-SDR. The darker bars with vertical hatches correspond to the $\Delta$SI-SDR.}
	\label{fig:under}
\end{figure}

We now focus on the performance of the proposed methods in the case where the number of nodes is smaller than the number of sources. This could happen in a real situation, for example if a phone shuts down during the meeting. The performance for the scenario with \mbox{$N=3$} sources is presented in Figure~\ref{fig:under} where we recall the performance of the equally-determined case (\mbox{$N=K=2$}) for an easier comparison.

In this under-determined case, even if there is only one more source than nodes, the multi-node solution performs worse than the two other ones. The mismatch between training and testing leads to worse performance. Since the single-node solution, as well as the \ac{mwf}, performs quite well, it means that the drop of performance of the multi-node solution is due to the multi-node \ac{crnn} that is trained on mixtures with only one interferent speaker while tested on mixtures with two interefent speakers. A similar behaviour was observed in the under-determined cases with \mbox{$N=4$} sources.

This indicates that dealing with under-determined cases probably requires to train specific networks with the proper number of interferent sources. Training and testing on a variable number of sources remains an open challenge. This could be addressed within our spatially distributed framework by adapting a dedicated strategy, e.g. in estimating the sources iteratively \cite{Kinoshita2018} or by adapting the loss function \cite{Luo2020, Turpault2020}.

\section{Conclusion}\label{sec:conclusion}
We introduced a distributed algorithm that can process spatial information captured by a spatially unconstrained microphone array. The spatial information is propagated over the microphone array to deliver to all the nodes a global insight into the whole acoustic scene. We evaluated the algorithm in typical meeting configurations and showed that thanks to the spatial information, a CRNN can predict accurate TF masks which lead to almost oracle performance. In scenarios where the number of nodes match the number of sources, we showed that the performance increases when the number of sources (and nodes) increases. We also analysed the limits of this approach when the number of nodes does not match the number of sources. Solving this problem could require the adaptation of techniques developped in the case of fixed microphone arrays.
\vfill\pagebreak

\bibliographystyle{IEEEbib}
\bibliography{strings,refs}

\end{document}